\documentclass{article}

\usepackage[final]{neurips_2023}

\usepackage[utf8]{inputenc} % allow utf-8 input
\usepackage[T1]{fontenc}    % use 8-bit T1 fonts
\usepackage{hyperref}       % hyperlinks
\usepackage{url}            % simple URL typesetting
\usepackage{booktabs}       % professional-quality tables
\usepackage{amsfonts}       % blackboard math symbols
\usepackage{nicefrac}       % compact symbols for 1/2, etc.
\usepackage{microtype}      % microtypography
\usepackage{xcolor}         % colors

\usepackage{colortbl}
\usepackage{graphicx}
\usepackage{subcaption}
\usepackage{amsmath}

\title{Towards Joint Sequence-Structure Generation\\ of Nucleic Acid and Protein Complexes\\ with $\mathrm{SE(3)}$-Discrete Diffusion
}

\author{%
  Alex Morehead\thanks{Alex Morehead was employed by Profluent Bio as a research intern at the time of this work.} , Jeffrey A. Ruffolo, Aadyot Bhatnagar, \& \textbf{Ali Madani} \\
  AI Research Lab \\
  Profluent Bio \\
  Berkeley, CA, USA \\
  \texttt{acmwhb@missouri.edu,\ \{jruffolo, abhatnagar, ali\}@profluent.bio} \\
}

\begin{document}

\maketitle

\begin{abstract}
    % Template ML abstract template (no need to go with this): 
    % First sentence is the importance of the work.
    % Second sentence is a problem that still exists as barrier to its success.
    % Third sentence is how we addressed the problem high level
    % Fourth can provide some detail
    % Fifth highlights the empirical results
    % Sixth is an epic conclusion with mention of open sourcing.
    Generative models of macromolecules carry abundant and impactful implications for industrial and biomedical efforts in protein engineering. However, existing methods are currently limited to modeling protein structures or sequences, independently or jointly, without regard to the interactions that commonly occur between proteins and other macromolecules. In this work, we introduce \textsc{MMDiff}, a generative model that jointly designs sequences and structures of nucleic acid and protein complexes, independently or in complex, using joint $\mathrm{SE(3)}$-discrete diffusion noise. Such a model has important implications for emerging areas of macromolecular design including structure-based transcription factor design and design of noncoding RNA sequences. We demonstrate the utility of \textsc{MMDiff} through a rigorous new design benchmark for macromolecular complex generation that we introduce in this work. Our results demonstrate that \textsc{MMDiff} is able to successfully generate micro-RNA and single-stranded DNA molecules while being modestly capable of joint modeling DNA and RNA molecules in interaction with multi-chain protein complexes. Source code: \href{https://github.com/Profluent-Internships/MMDiff}{https://github.com/Profluent-Internships/MMDiff}.
\end{abstract}

\section{Introduction}
The ability to effectively design proteins using computational methods is a long-sought-after goal of protein engineering research \citep{LUTZ2010734}. Such techniques could enable accelerated development of vaccines targeting novel viruses, enhanced exploration of the space of designable materials, and the ability to engineer new protein-based energy resources towards a more sustainable climate future. Recent progress in computationally designing protein variants given an initial wildtype sequence has demonstrated the utility of machine learning in protein engineering \citep{HIE2022145}. More recently, deep learning models based on structure prediction networks have demonstrated the ability to flexibly design novel protein structures with desired functions in wet-lab experiments \citep{watson2023novo}. Nonetheless, considerably less attention has been directed toward the design of novel protein-nucleic acid interactions, which carry implications for computationally engineering new means of transcription factors and intervening in a host of essential functions within living organisms. Approaching such design tasks is arguably more challenging, as the number of high-quality macromolecular complexes currently available in the RCSB Protein Data Bank (PDB) \citep{berman2000protein} is notably scarce while the shapes of such complexes can also be much more flexible compared to proteins.

Towards this end, in this work we explore the design of new nucleic acid (NA) and protein complexes using a generative diffusion process we refer to as $\mathrm{SE(3)}$-discrete diffusion. In Section \ref{section:joint_diffusion}, we discuss the $\mathrm{SE(3)}$-discrete process which guides deep learning models to jointly reverse an $\mathrm{SE(3)}$ diffusion process \footnote{$\mathrm{SE(3)}$ refers to the group of 3D rotations and translations (excluding reflections), whereas $\mathrm{SE(3)}^{N}$ represents the manifold of $N$ $\mathrm{SE(3)}$ frames.} on rigid body frames in $\mathbb{R}^{3}$ \citep{yim2023se} and a related diffusion process on discrete sequence inputs \citep{chen2022analog}. In Section \ref{section:macromolecule_generation}, we introduce our proposed $\mathrm{SE(3)}$-discrete diffusion method for protein-nucleic acid design, \textsc{MMDiff}, which is to our best knowledge the \underline{first} generative diffusion model for the design of macromolecular sequences and structures beyond proteins. Through our experiments in Section \ref{section:experiments}, we introduce a systematic open-source suite of computational benchmarks to assess the ability of a generative macromolecular design model to produce designable, diverse, and novel samples, and, under such a framework, highlight the utility of \textsc{MMDiff} for macromolecular complex generation. Consequently, our contributions presented in this work pave the way for the development of future macromolecular generative models.

\begin{figure}[t]
\centering
\includegraphics[width=\linewidth]{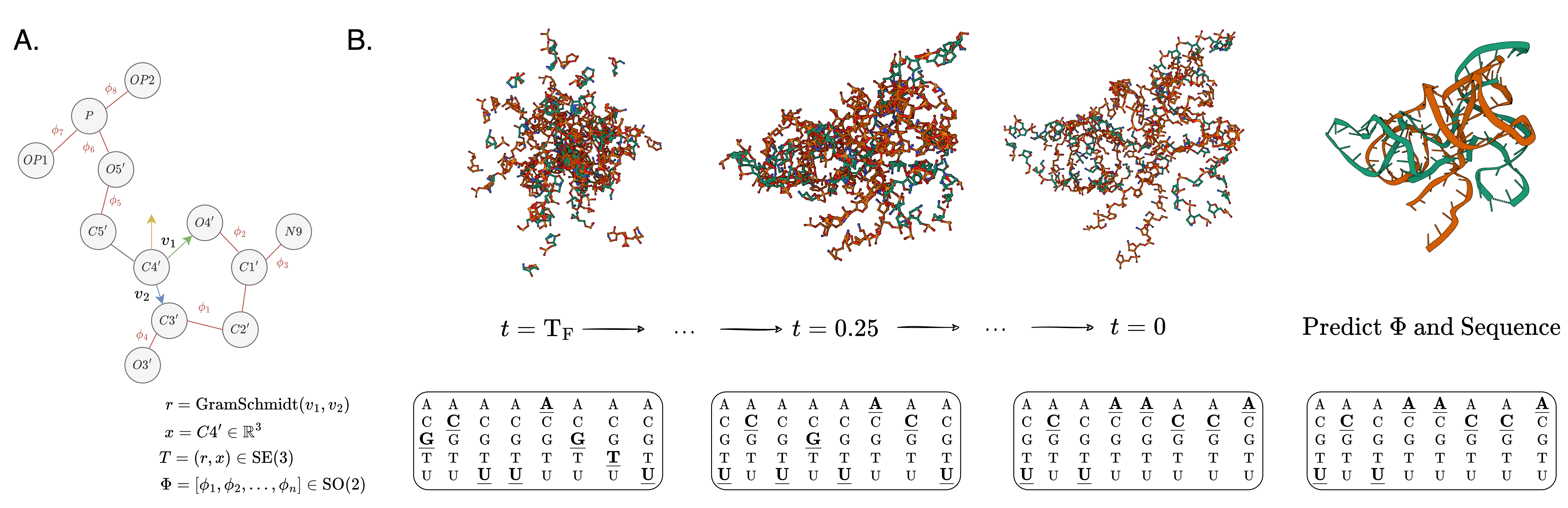}
\caption{Our proposed Macromolecular Diffusion Model (\textsc{MMDiff}) jointly designs macromolecular sequences and structures. \textbf{A.} An overview of \textsc{MMDiff}. For each nucleic acid residue, a rigid body frame centered at the $\mathrm{C4^{\prime}}$ atom is constructed. To build such a frame, the $\mathrm{GramSchmidt}$ algorithm is applied to a residue's $v_{1}$ and $v_{2}$ vectors, in the process placing its $C3^{\prime}$, $O4^{\prime}$, and $C5^{\prime}$ atoms with respect to the position of the $C4^{\prime}$ atom. The positions of all other residue atoms are placed autoregressively according to a corresponding torsion angle ($\Phi$) predicted by \textsc{MMDiff}. \textbf{B.} An illustrative example of how \textsc{MMDiff} generates realistic macromolecular samples. Through the iterative process of denoising $N$ geometric frames and $N$ one-hot sequence vectors initialized from their respective reference distributions, \textsc{MMDiff} transitions an initially-random sequence-structure pair at timestep $\mathrm{T_{F}}$ into a coherent macromolecule at timestep $0$, at which point each frame and its associated torsion angles are used to construct the position of each atom.}
\label{fig:MMDiff}
\end{figure}

\section{Methodology}
\subsection{Preliminaries and Notation}
\label{section:preliminaries}
\textbf{Protein geometry parametrization.} Inspired by the recent work of \cite{yim2023se}, we adopt the standard backbone parametrization of AlphaFold 2 for protein geometry modeling. Succinctly, with such a parametrization, $N_{p}$ protein residues are modeled as $N_{p}$ rigid body frames centered at each residue's $\mathrm{C^{\star}_{\alpha}}$ atom ($\mathrm{C^{\star}_{\alpha}} = (0, 0, 0)$), with each frame representing a residue's $\mathrm{N}^{\star}, \mathrm{C^{\star}_{\alpha}}, \mathrm{C}^{\star}$, and $\mathrm{O}^{\star}$ (local) atoms, respectively. For a given protein residue indexed by $n$, to obtain the absolute (global) positions of its main atoms, one may apply a special Euclidean ($\mathrm{SE(3)}$) transformation $T_{n}$ as
\begin{equation}
    [\mathrm{N}_{n}, \mathrm{C}_{n}, (\mathrm{C_{\alpha}})_{n}] = T_{n} \cdot [\mathrm{N}^{\star}, \mathrm{C}^{\star}, \mathrm{C^{\star}_{\alpha}}],
\end{equation}
where $T_{n} = (r_{n}, x_{n})$ is composed of two components, $r_{n} \in \mathrm{SO(3)}$ being a $3 \times 3$ rotation matrix and $x_{n} \in \mathbb{R}^{3}$ being a translation vector. Subsequently, all $N_{p}$ frames are denoted as $\mathbf{T_{p}} = [T_{1}, \dots, T_{N_{p}}] \in \mathrm{SE(3)}^{N_{p}}$. Lastly, one may place each residue's $\mathrm{O}^{\star}$ atom by rotating it around the corresponding residue's $C_{\alpha}-C$ bond using the torsion angle $\psi$. For additional details regarding the mapping between idealized and absolute protein atom positions, we refer readers to \cite{yim2023se}.

\textbf{Nucleic acid geometry parametrization.} To model nucleic acid residues jointly with protein residues, we require an expressive frame encoding of nucleic acid geometry. Towards this end, we propose to model $N_{n}$ nucleic acid residues as $N_{n}$ rigid body frames centered at each residue's $\mathrm{C4^{\prime}}^{\star}$ atom ($\mathrm{C4^{\prime}}^{\star} = (0, 0, 0)$), with each frame representing a residue's $\mathrm{C1^{\prime}}^{\star}$ through $\mathrm{C5^{\prime}}^{\star}$ atoms, $\mathrm{O3^{\prime}}^{\star}$ through $\mathrm{O5^{\prime}}^{\star}$ atoms, as well as its $\mathrm{P}^{\star}$, $\mathrm{OP1}^{\star}$, $\mathrm{OP2}^{\star}$, and $\mathrm{N9}^{\star}$ (or, for pyrimidine residues, $\mathrm{N1}^{\star}$) atoms, respectively. Then for a given nucleic acid residue indexed by $n$, to obtain the absolute positions of its main atoms, one may apply an $\mathrm{SE(3)}$ transformation $T_{n} = (r_{n}, x_{n})$ as
\begin{equation}
    [\mathrm{C3^{\prime}}_{n}, \mathrm{O4^{\prime}}_{n}, \mathrm{C4^{\prime}}_{n}, \mathrm{C5^{\prime}}_{n}] = T_{n} \cdot [\mathrm{C3^{\prime}}^{\star}, \mathrm{O4^{\prime}}^{\star}, \mathrm{C4^{\prime}}^{\star}, \mathrm{C5^{\prime}}^{\star}],
\end{equation}
where all $N_{n}$ frames are denoted as $\mathbf{T_{n}} = [T_{N_{p} + 1}, \dots, T_{N_{n}}] \in \mathrm{SE(3)}^{N_{n}}$. For nucleic acid residues, to place each residue's auxiliary atoms (i.e., $\mathrm{C2^{\prime}}^{\star}_{n}, \mathrm{C1^{\prime}}^{\star}_{n}, \mathrm{N9}^{\star}_{n} (\mbox{or } \mathrm{N1}^{\star}_{n}), \mathrm{O3^{\prime}}^{\star}_{n}, \mathrm{O5^{\prime}}^{\star}_{n}, \mathrm{P}^{\star}_{n}, \mathrm{OP1}^{\star}_{n}, \mbox{and } \mathrm{OP2}^{\star}_{n})$, one may autoregressively rotate them around the corresponding residue's $\mathrm{C4^{\prime}}-\mathrm{C3^{\prime}}$, $\mathrm{C4^{\prime}}-\mathrm{O4^{\prime}}$, $\mathrm{O4^{\prime}}-\mathrm{C1^{\prime}}$, $\mathrm{C4^{\prime}}-\mathrm{C3^{\prime}}$, $\mathrm{C4^{\prime}}-\mathrm{C5^{\prime}}$, $\mathrm{C5^{\prime}}-\mathrm{O5^{\prime}}$, $\mathrm{O5^{\prime}}-\mathrm{P^{\prime}}$, and $\mathrm{O5^{\prime}}-\mathrm{P^{\prime}}$ bonds using the torsion angles $\Phi = [\phi_{1},\dots,\phi_{8}]$, respectively. For additional background regarding nucleic acid geometry, we refer readers to \cite{gelbin1996geometric}.

\textbf{Joint protein-nucleic acid geometry parametrization.} Given $N_{p}$ protein frames and $N_{n}$ nucleic acid frames, respectively, we subsequently denote the manifold of frames that we aim to represent with a generative model as $\mathbf{N} = (N_{p}, N_{n})$. Consequently, we recover the original means of modeling a distribution $\mathbf{X}^{(0)} \sim p_{0}$ of frames in $\mathrm{SE(3)}^{\mathbf{N}}$ supported on a Riemannian manifold $\mathcal{M}$, as proposed by \cite{yim2023se}. In short, using a score network $s_{\theta}(t, \cdot)$, this allows one to model $\mathbf{X}^{(0)}$ via an approximation of the Stein score $\nabla \log p_{t}$ using the $\mathrm{SE(3)}$ denoising score matching (DSM) loss \citep{yim2023se}:
\begin{equation}
\label{eq:dsm_loss}
    \mathcal{L}(\theta) = \mathbb{E}[\lambda_{t} \lVert \nabla \log p_{t | 0}(\mathbf{X}^{(t)} | \mathbf{X}^{(0)}) - s_{\theta}(t, \mathbf{X}^{(t)}) \rVert^{2}],
\end{equation}
where $\theta$ are the network parameters to be optimized by minimizing $\mathcal{L}$; $p_{t | 0}$ is the density of $\mathbf{X}^{(t)}$ given $\mathbf{X}^{(0)}$; $\lambda_{t} > 0$ is a loss weight; and the expectation $\mathbb{E}$ is taken over $(\mathbf{X}^{(0)}, \mathbf{X}^{(t)})$ and $t \sim \mathcal{U}([0, \mathrm{T_{F}}])$, with $\mathrm{T_{F}}$ being the final timestep in a diffusion (i.e., noising) process over $\mathbf{X}^{(0)}$.

\textbf{Remaining preliminaries.} Following the notation of \cite{yim2023se}, bold denotes the concatenation of variables (e.g., $\mathbf{n} = (n_{1}, \dots, n_{N})$); uppercase denotes stochastic variables (e.g., $X \sim p$), whereas lowercase is reserved for deterministic variables; and superscripts with parentheses denote time (e.g., $x^{(t)}$).

\subsection{Joint Continuous-Discrete Diffusion in $\mathbb{R}^{3}$}
\label{section:joint_diffusion}
\textbf{Sequence modeling.} Expressively modeling arbitrary macromolecular structures is a non-trivial task for a generative model to address. Nonetheless, in this work, we argue that an understanding of macromolecules is inherently incomplete without also considering the sequence space of such molecules. In particular, modeling of macromolecular sequences enables methods to learn molecular representations that are complementary, and in many cases foundational, to their structure knowledge of 3D molecular structures \citep{jumper2021highly, lin2022language, baek2022accurate, shen2022e2efold}. This is particularly relevant for nucleic acid modeling in that, practically speaking, no general-purpose existing methods currently exist for fixed-backbone sequence design of nucleic acid structures, although RNA-specific sequence design methods are beginning to emerge \cite{joshi2023multi}. As such, in this work, we are also interested in modeling the design space of macromolecular sequences. We do so by introducing a discrete diffusion framework that, with key adaptations to its noise scheduling, is amenable to the SE(3) Diffusion framework of \cite{yim2023se} described above in Section \ref{section:preliminaries}. This allows our proposed method, \textsc{MMDiff}, to jointly generate sequence-structure pairs of new macromolecules, thereby circumventing the need for fixed-backbone sequence design methods for nucleic acid residues given a generated backbone structure.

\textbf{Discrete sequence generation.} Diffusion generative modeling of discrete data types such as amino acid and nucleic acid sequences has received considerably less attention in the research literature compared to diffusion generative approaches to modeling continuous data such as images or even 3D point cloud coordinates. Nonetheless, preliminary works have investigated alternative approaches to expressively modeling discrete values in a diffusion generative setting \citep{hoogeboom2021argmax, NEURIPS2021_958c5305, chen2022analog}, with some directly adding discrete noise to one's discrete network inputs and others first representing discrete data in a continuous space and then applying standard Gaussian diffusion techniques to generate new discrete data samples through the use of an \textsc{argmax} operation \citep{ho2020denoising}. In this work, we adopt the latter approach of modeling discrete macromolecular sequence data as continuous representations such that (1) methodological developments for continuous diffusion models will be directly amenable to such an approach and (2) it then becomes trivial to condition sample generation on arbitrary sequence features of interest (e.g., secondary structure annotations for protein residues) by embedding such features in the same continuous space as the input sequence. Concretely, inspired by \cite{lisanza2023joint}, we propose to model a discrete amino acid and nucleic acid sequence $\mathbf{S}_{d}$ in a shared vocabulary space (i.e., $\mathbb{R}^{s \times n}$) as a zero-centered continuous one-hot vector representation
\begin{equation}
    \mathbf{S}_{c}^{(0)} = 2 * \mathrm{onehot}(\mathbf{S}_{d}) - 1 = \{2 * (s \in \{0.0, 1.0\}^{n} : \sum_{i = 1}^{n} s_{i} = 1.0) - 1, \forall s \in \mathbf{S}_{d})\},
\end{equation}
such that discrete sequences can be generated by reversing a continuous Gaussian diffusion process
\begin{equation}
\label{eq:sequence_diffusion_process}
    q(\mathbf{S}_{c}^{(t_{d})} | \mathbf{S}_{c}^{(0_{d})}) = \mathcal{N}(\mathbf{S}_{c}^{(t_{d})} | \sqrt{\bar{\alpha}^{(t_{d})}} \mathbf{S}_{c}^{(0_{d})}, (1 - \bar{\alpha}^{(t_{d})}) \mathrm{I})
\end{equation}
on an initially-random sequence vector representation $\mathbf{S}_{c}^{(T_{d})}$ corresponding to the final sequence diffusion timestep $T_{d}$. Here, $\bar{\alpha}^{(t_{d})}$ represents $1 - \beta^{(t_{d})}$, where $\beta^{(t_{d})}$ denotes one's chosen variance schedule for the sequence diffusion process, and $t_{d}$ represents a discretized version of the structure diffusion continuous timestep $t$ via equally-spaced binning of such timesteps. Note that this binning operation is important to ensure that the sequence diffusion and structure diffusion processes are in alignment with one another at each step in their respective diffusion trajectories.

\textbf{Training and sampling with structure noise.} Reversal of the structure noise schedule is performed by first using the FrameDiff algorithm of \cite{yim2023se} to predict at each structure diffusion timestep the denoised frames $\mathbf{F}^{(0)}$ to derive the respective rotation ($r$) and translation ($x$) scores $\{(\varsigma^{r}_{\theta, n}, \varsigma^{x}_{\theta, n})\}_{n = 1}^{\mathbf{N}}$, where during model training such score predictions are supervised using Equation \ref{eq:dsm_loss} along with supervision for predicted torsions $\Phi$ through auxiliary backbone atom and inter-atom distance losses. The scores $\{(\varsigma^{r}_{\theta, n}, \varsigma^{x}_{\theta, n})\}_{n = 1}^{\mathbf{N}}$ are then used to transition to a previous structure diffusion timestep as $\mathbf{F}^{(t - 1)} = \{\exp_{\mathbf{F}_{n}^{(t)}}\{(W_{n}^{r}, W_{n}^{x})\}\}_{n = 1}^{\mathbf{N}}$, where $(W_{n}^{r}, W_{n}^{x})$ are derived independently as $W_{n}^{x} = P[P\gamma[\frac{1}{2} X_{n}^{(t)} + \varsigma_{\theta, n}^{x}] + \zeta \sqrt{\gamma} [Z_{n}^{x} \sim \mathcal{N}(0, \mathrm{Id}_{3})]]$ and $W_{n}^{r} = \gamma \varsigma_{\theta, n}^{r} + \zeta \sqrt{\gamma} [Z_{n}^{r} \sim \mathcal{T} \mathcal{N}_{R_{n}^{(t)}}(0, \mathrm{Id})]$, respectively \citep{yim2023se}. Here, $P \in \mathbb{R}^{3\mathbf{N} \times 3\mathbf{N}}$ represents a projection matrix that removes an object's center of mass; $\gamma = \frac{1 - \epsilon}{N_{\mathrm{steps}}}$ with $\epsilon$ being the final continuous timestep of the structure diffusion noise schedule prior to predicting torsions $\Phi$; $X_{n}^{(t)} \in \mathbb{R}^{3}$ and $R_{n}^{(t)} \in \mathbb{R}^{3 \times 3}$ denote the translation vector and rotation matrix associated with frame (i.e., residue) $\mathrm{F}_{n}^{(t)}$ at continuous timestep $t$; $Z_{n}^{x}$ and $Z_{n}^{r}$ denote Gaussian noise sampled from the tangent spaces of $X_{n}^{(t)}$ and $R_{n}^{(t)}$, respectively; and $\zeta \in [0, 1]$ represents a scaling factor for structure diffusion noise.

\textbf{Training and sampling with sequence noise.} As pointed out by \cite{li2022diffusion, lisanza2023joint}, in the context of training diffusion models for discrete sequence generation it is also desirable to use a square root diffusion noise schedule for $\beta^{(t_{d})}$, as using other diffusion noise schedules during training may make it trivial for a model to predict a ground-truth sequence from a small-noise discrete timestep $t_{d}$ using the \textsc{argmax} operation. Such prior works thereby sensitize the model to sequence noise at small $t_{d}$ timesteps by training it with a square root noise schedule, while supervising its sequence predictions using a categorical cross-entropy loss. After model training with a square root noise schedule, in this work we also explore sampling sequences from a trained model using cosine and linear noise schedules instead, to investigate how doing so affects the model at inference time with regards to its sequence generation trajectories (and thereby its structure generation trajectories), especially at small $t_{d}$ timesteps. For interested readers, these additional experiments are presented in Appendix \ref{appendix:alternative_sequence_noise}. Reversal of the chosen sequence noise schedule is then performed by having a trained model directly predict $\mathbf{S}_{c}^{(0_{d})}$ at timestep $t_{d}$ and then using this value in Equation \ref{eq:sequence_diffusion_process} to transition to timestep $t_{d} - 1$.

\subsection{Generating macromolecular complexes with \textsc{MMDiff}}
\label{section:macromolecule_generation}

As illustrated in Figure \ref{fig:MMDiff}, our proposed method, \textsc{MMDiff}, enables joint generation of nucleic acid sequences and structures through a hybrid $\mathrm{SE(3)}$-discrete diffusion process. Moreover, by way of its design as described in Section \ref{section:preliminaries}, \textsc{MMDiff} can simultaneously model protein sequences and structures together with nucleic acids, enabling the generation of inter-molecular complexes. In the remainder of this section, we will discuss the remaining details necessary to train and sample from \textsc{MMDiff} for generating macromolecular complexes.

\textbf{Model architecture, data, and featurization.} Inspired by its demonstrated ability to generate designable, diverse, and novel monomeric protein structures, for \textsc{MMDiff} we adapt the FrameDiff model architecture of \cite{yim2023se} for modeling of macromolecular complexes. To train \textsc{MMDiff} for macromolecular modeling, we downloaded from the RCSB PDB \citep{berman2000protein} all available instances of protein-nucleic acid structures using the \textsc{PDBManager} API within the \textsc{Graphein} Python package \citep{jamasb2022graphein}. We filtered the downloaded structures to those that (1) were determined at a resolution of 4.5 \r{A} or better using nuclear magnetic resonance, diffraction, or cryo-electron microscopy; (2) contain protein chains between lengths 40 and 256 or nucleic acid chains between lengths 3 and 256; and (3) contain no more than 10 protein or 10 nucleic acid chains. In Appendix \ref{appendix:dataset_distributions}, we plot the length and chain composition distributions of the various training dataset configurations we explore in this work. To accurately characterize the intricate chain compositions present within many macromolecular PDB complexes, we subsequently adopt the relative position encoding of AlphaFold-Multimer \citep{evans2021protein} in contrast to the absolute position encoding used by \cite{yim2023se}. Otherwise, we adapt the feature initialization scheme of \cite{trippe2022diffusion} to accommodate both proteins and nucleic acids.

\textbf{Intra-molecule consensus sampling.} To generate internally-consistent macromolecular sequences, we propose a consensus sampling algorithm that, during sampling, constrains \textsc{MMDiff} to select only amino acid residue types in a designated protein sequence region and only the residues types of a single type of nucleic acid molecule (i.e., either DNA or RNA) in a nucleic acid sequence region. Namely, for a given sequence representation $\mathbf{S}_{c}^{(t_{d})}$, we propose to consensus-sample sequences by dynamically modifying the pseudo-probabilities (i.e., continuous representation) of $\mathbf{S}_{c}^{(t_{d})}$ such that the largest probabilities associated with residue types outside of the current sequence's molecule type consensus (e.g., $C^{(t_{d})} = [\textsc{DA}, \textsc{DC}, \textsc{DG}, \textsc{DT}] = $ DNA residue types if 50\% or more of the generated residues are DNA residues) are set to be marginally smaller than the smallest probabilities associated with residue types within the current molecule type consensus as
\begin{equation}
\label{eq:consensus_sampling}
    \mathbf{S}_{c}^{(t_{d})} = \{s \in \mathbb{R}^{n} : \max_{i \notin C^{(t_{d})}} s_{i} < \min_{i \in C^{(t_{d})}} s_{i},\ \forall s \in \mathbf{S}_{c} \mbox{ with } i \in [1, n]\}.
\end{equation}
Applying Equation \ref{eq:consensus_sampling} after each iteration of sequence denoising during sampling effectively biases the generative sequence distribution from which \textsc{MMDiff} samples new sequences, such that the model is increasingly less likely to naturally sample out-of-vocabulary within a given chain.

\section{Experiments}
\label{section:experiments}
\begin{table}[!htb]
    \centering
    \begin{subtable}{\textwidth}
        \centering
        \begin{tabular}{ccccccc}
             \toprule
             Method & $< 5 \text{\r{A}}\ \mathrm{scRMSD}$ & $\mathrm{D}_{chain}$ & $\mathrm{D}_{complex}$ & $\mathrm{D}_{single}$ & $\mathrm{D}_{all}$ & $\mathrm{N}_{pool}$ \\
             \midrule
             Random Generation & 0.00\% & 0.92 & 1.00 & 1.00 & 1.00 & 0.85 \\
             \rowcolor[gray]{0.8} \textsc{MMDiff} & \textbf{0.74}\% & \textbf{0.92} & \textbf{1.00} & \textbf{1.00} & \textbf{1.00} & 0.72  \\
             \rowcolor[gray]{0.8} \textsc{MMDiff-Protein} & 0.00\% & 0.91 & 0.99 & 0.99 & 0.99 & 0.73  \\
             \bottomrule
        \end{tabular}
        \caption{Protein Generation}
        \label{subtab:protein_generation}
    \end{subtable}
    
    \vspace{1em}
    
    \begin{subtable}{\textwidth}
        \centering
        \begin{tabular}{ccccccc}
             \toprule
             Method & $< 5 \text{\r{A}}\ \mathrm{scRMSD}$ & $\mathrm{D}_{chain}$ & $\mathrm{D}_{complex}$ & $\mathrm{D}_{single}$ & $\mathrm{D}_{all}$ & $\mathrm{N}_{pool}$ \\
             \midrule
             Random Generation & 1.33\% & 0.99 & 1.00 & 0.99 & 0.99 & 0.75 \\
             \rowcolor[gray]{0.8} \textsc{MMDiff} & \textbf{8.67}\% & \textbf{1.00} & \textbf{1.00} & \textbf{1.00} & \textbf{1.00} & 0.74 \\
             \rowcolor[gray]{0.8} \textsc{MMDiff-NA} & 6.00\% & 0.98 & 0.98 & 0.99 & 0.99 & 0.65 \\
             \bottomrule
        \end{tabular}
        \caption{Nucleic Acid Generation}
        \label{subtab:na_generation}
    \end{subtable}
    
    \vspace{1em}
    
    \begin{subtable}{\textwidth}
        \centering
        \begin{tabular}{ccccccc}
             \toprule
             Method & $< 5 \text{\r{A}}\ \mathrm{scRMSD}$ & $\mathrm{D}_{chain}$ & $\mathrm{D}_{complex}$ & $\mathrm{D}_{single}$ & $\mathrm{D}_{all}$ & $\mathrm{N}_{pool}$ \\
             \midrule
             Random Generation & 0.37\% & 1.00 & 1.00 & 1.00 & 1.00 & 0.72 \\
             \rowcolor[gray]{0.8} \textsc{MMDiff} & \textbf{0.74}\% & 0.96 & 0.85 & \textbf{1.00} & \textbf{1.00} & 0.69 \\
             \rowcolor[gray]{0.8} \textsc{MMDiff-Monomer} & 0.00\% & 0.96 & 0.85 & 1.00 & 1.00 & 0.70 \\
             \bottomrule
        \end{tabular}
        \caption{Protein-Nucleic Acid Complex Generation}
        \label{subtab:protein_na_complex_generation}
    \end{subtable}
    
    \caption{A comparison of our proposed method, \textsc{MMDiff}, against a random baseline for generating different types of biological complexes, where we assess each method's designability (based on an $\mathrm{scRMSD}$ threshold), diversity (regarding four variants of $\mathrm{D}$), and novelty (i.e., $\mathrm{N}_{pool}$). Across all tasks, \textsc{MMDiff} represents the same model trained on protein-nucleic acid complexes. However, in task (a), \textsc{MMDiff-Protein} is trained on both protein monomers and protein complexes; in task (b), \textsc{MMDiff-NA} is trained solely on nucleic acid complexes; and lastly, in task (c), \textsc{MMDiff-Monomer} is trained on both monomers and protein-nucleic acid complexes.}
    \label{tab:combined_complex_generation_results}
\end{table}
\textbf{Metrics.} To investigate the performance of \textsc{MMDiff} (and similar generative models developed in the future) in generating new macromolecular complexes, we propose to examine the quality of its generated complexes from multiple perspectives: (1) the model's ability to generate \textit{designable} sequences such that, upon predicting the structure of such a designed sequence using an external structure prediction model, the difference between the model's co-designed structure and the sequence-predicted structure is minimal; (2) the model's ability to generate \textit{diverse} complex structures that are distinct from each other after structure-based clustering; and (3) the model's ability to generate (3) \textit{novel} complex structures that are well-differentiated from complex structures found within the model's training dataset.

\textbf{Designability.} To instantiate such metrics in the context of macromolecular complex generation as outlined above, for (1) we adopt RoseTTAFold2NA \citep{baek2022accurate} for sequence-based prediction of generated protein-nucleic acid structures to assess \textit{designability}, where we report structural similarities in terms of self-consistency RMSD and TM-score (i.e., $\mathrm{scRMSD}$ and $\mathrm{scTM}$) for each complex as determined by US-align \citep{zhang2022us}. Due to practical concerns regarding the significant computational requirements of running RoseTTAFold2NA for each complex, in this setting, we generate and evaluate one sequence per structure produced by the model, using RoseTTAFold2NA in \textit{single-sequence} mode to reduce its average prediction time from 25 minutes to 2 minutes. Note that this may limit the sequence designability performance that \textsc{MMDiff} can achieve, as previous works have found it necessary to design up to 8 sequences per generated structure to achieve good designability results. Moreover, RoseTTAFold2NA was trained to predict protein-nucleic acid structures using multiple sequence alignments (MSAs) as a crucial input feature, hence running the model in \textit{single-sequence} mode will incur some prediction accuracy degradation in exchange for practical running times. Taking all of these constraints into account, we propose to evaluate each method's ability to generate molecules with an $\mathrm{scRMSD}$ $< 5$ \r{A} to denote a successfully-designed macromolecule, with an $\mathrm{scRMSD}$ $< 2$ \r{A} being a more ideal threshold \citep{yim2023se} yet being largely impractical due to the potential mismatches between generated and predicted inter-molecule interaction types (e.g., RF2NA predicting non-interacting chain structures and \textsc{MMDiff} generating interacting chain structures).

\begin{figure}[!htb]
    \centering
    \includegraphics[width=\linewidth]{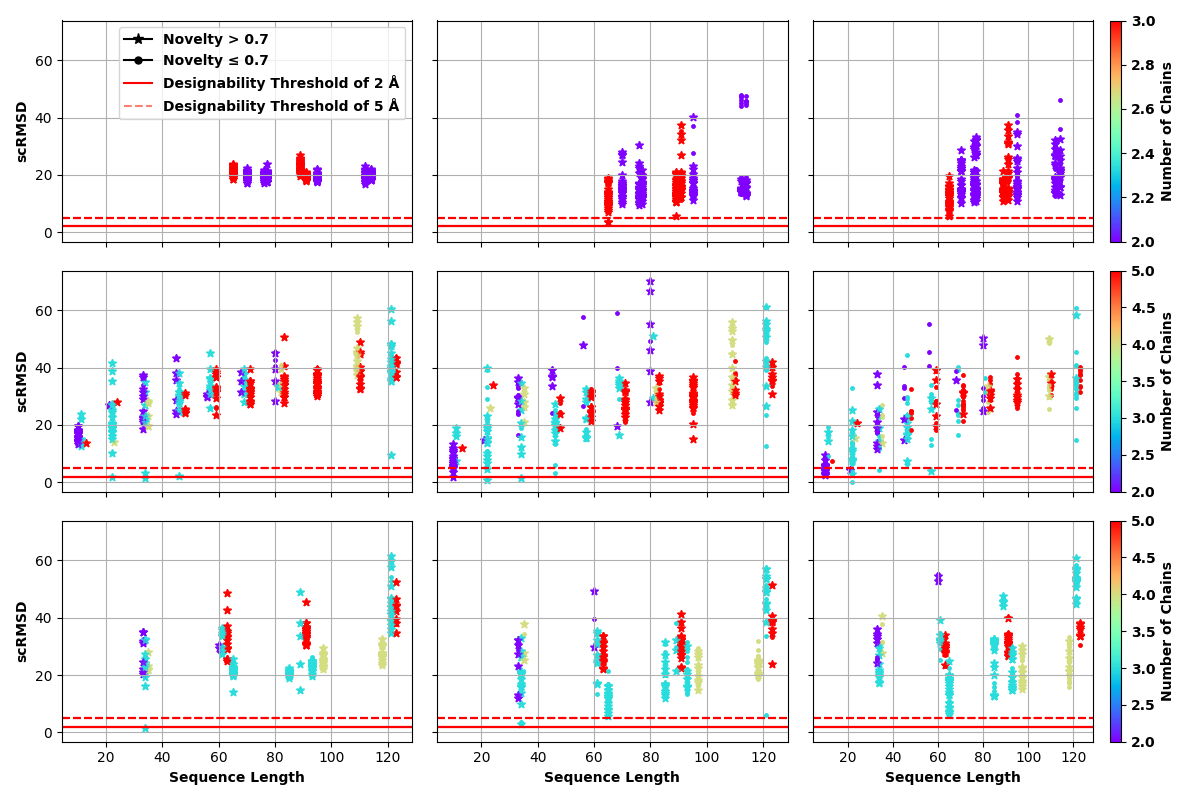}
    
    \caption{Comparison of $\mathrm{scRMSD}$ complex designability results using different training methods. Here, the top row corresponds to protein-only experiments, the middle row to nucleic acid-only experiments, and the bottom row to protein-nucleic acid experiments. The columns denotes samples generated using the random macromolecule generation baseline, \textsc{MMDiff}, and \textsc{MMDiff-\{Protein, NA, Monomer\}} (corresponding to the \{first, second, third\} row), respectively. Note that novel data samples are displayed with a \textsc{*} symbol. Overall, most designable complexes contain 2-3 chains, and most generated complexes contain novel chains (with a novelty $> 0.7$).}
    \label{fig:designability_scrmsd_results}
\end{figure}

\textbf{Diversity.} Regarding (2), we use qTMclust \citep{zhang2022us} for macromolecular structure-based clustering of the model's pool of generated complex structures to measure \textit{diversity}, where the diversity of a pool is defined in four components as follows. We define the baseline $\mathrm{chain\ diversity}$ ($\mathrm{D}_{chain}$) of a pool as: (number of chain clusters) / (number of chains). Similarly, as an upper bound on the model's diversity, we define the simpler $\mathrm{complex\ diversity}$ ($\mathrm{D}_{complex}$) of a pool as: (number of complex clusters) / (number of complexes). More specifically, to measure how often the model generates a unique interaction type, we then define the $\mathrm{single-chain\ complex\ diversity}$ ($\mathrm{D}_{single}$) as: (number of complexes for which one of its chains is a chain cluster representative) / (number of complexes). Lastly, as a lower bound on the model's diversity, we define the $\mathrm{all-chain\ complex\ diversity}$ ($\mathrm{D}_{all}$) as: (number of complexes for which each of its chains is a chain cluster representative) / (number of complexes).

\textbf{Novelty.} To address (3) we perform pairwise structural alignments between the model's pool of generated complex structures and the model's training dataset using US-align \citep{zhang2022us} to estimate \textit{novelty}, where the novelty of a pool is quantified as the inverse of the maximum TM-score between a generated complex structure and any complex structure in the training dataset (i.e., $\mathrm{N}_{pool} = 1 - \mathrm{maxTM}$).

\textbf{Baselines.} To the best of our knowledge, no prior methods can co-design sequences and structures of macromolecular complexes. As such, in addition to evaluating \textsc{MMDiff}, we report results using a naive baseline that randomly generates sequences and structures of protein, nucleic acid, and protein-nucleic acid complexes, respectively, as a simple baseline to evaluate a macromolecular generative model's performance. For the random baseline, we extract the initial sequence and structure noise representations of \textsc{MMDiff}'s diffusion process (i.e., noise representations at timestep $T$) as the random baseline's generated outputs. As such, one should expect a generative model that has learned meaningful sequence and structural patterns from its training dataset to outperform such a baseline regarding the designability of its samples by fully reversing its learned diffusion process.

\begin{figure}[t]
    \centering
    \begin{subfigure}[b]{0.32\linewidth}
        \includegraphics[width=\linewidth]{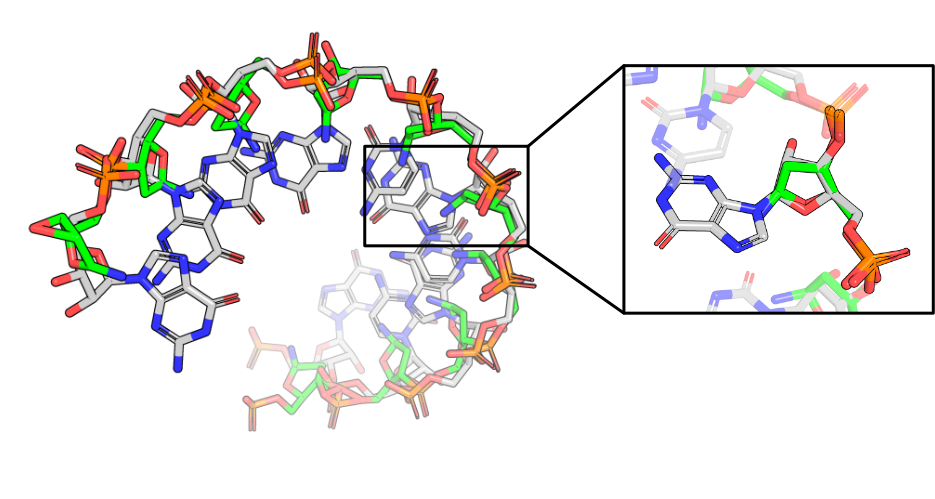}
        \includegraphics[width=\linewidth]{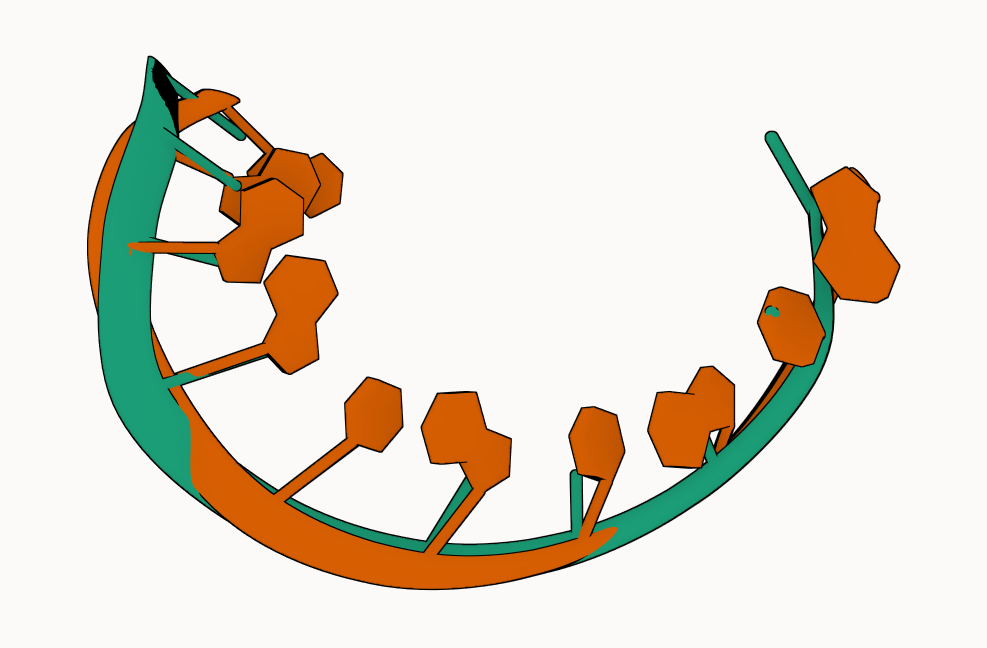}
        \caption{Designed micro-RNA.}
        \label{fig:designed_micro_rna}
    \end{subfigure}
    \begin{subfigure}[b]{0.32\linewidth}
        \includegraphics[width=\linewidth]{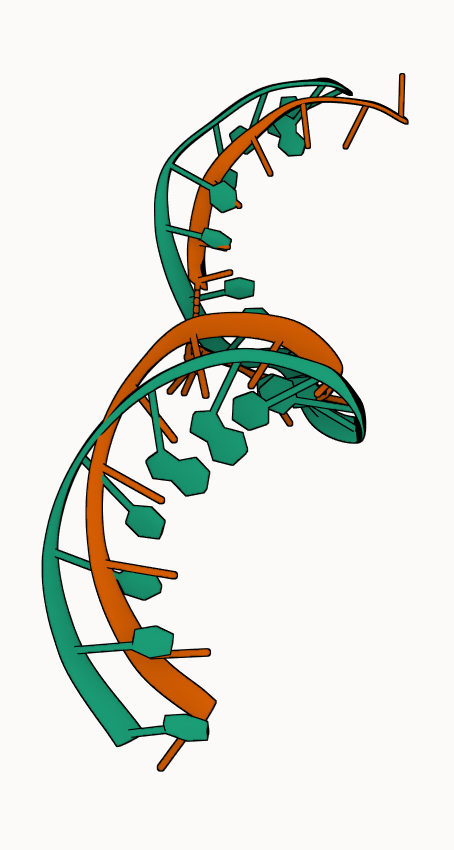}
        \caption{Designed single-stranded DNA.}
        \label{fig:designed_micro_s_dna}
    \end{subfigure}
    \begin{subfigure}[b]{0.32\linewidth}
        \includegraphics[width=\linewidth]{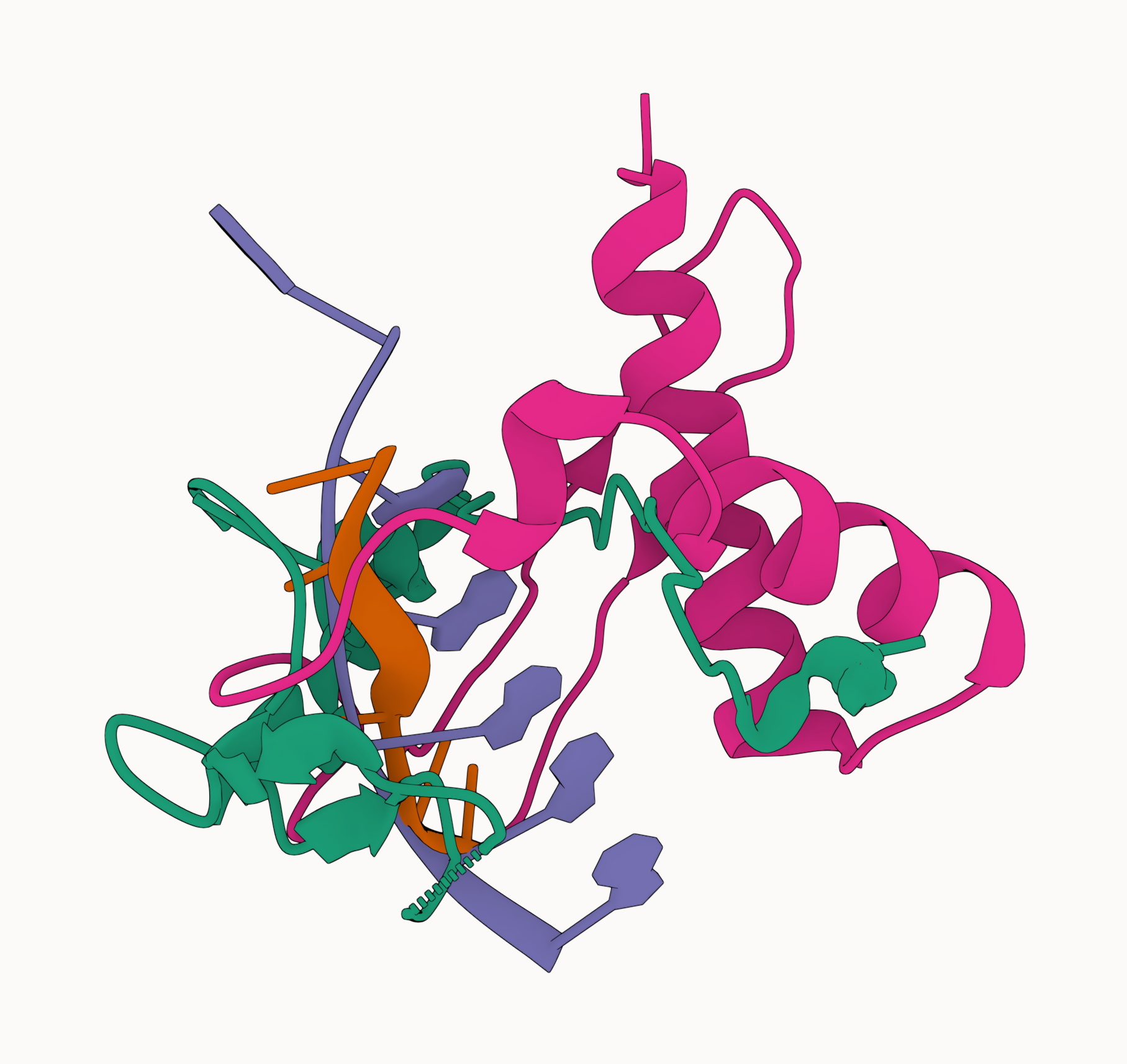}
        \caption{Designed protein-DNA.}
        \label{fig:designed_protein_dna_complex}
    \end{subfigure}
    \caption{Examples of macromolecules successfully designed by \textsc{MMDiff}.}
    \label{fig:combined_examples}
\end{figure}

\textbf{Results and Analysis.} As shown in Table \ref{subtab:protein_generation} and Figure \ref{fig:designability_scrmsd_results}, the results for protein complex generation suggest that \textsc{MMDiff} (trained on protein-nucleic acid complexes) can successfully generate a handful of designable protein complexes, whereas the random baseline cannot. Interestingly, \textsc{MMDiff-Protein}, a version of \textsc{MMDiff} that is trained not only on protein-nucleic acid complexes but also jointly on the large monomeric protein structure dataset of \cite{yim2023se}, underperforms \textsc{MMDiff} in all design metrics. One explanation for this observation is that the generative modeling space that \textsc{MMDiff-Protein} is tasked to model when jointly training on protein monomers is much larger and noisier compared to the modeling space of solely protein-nucleic acid complexes, a notion that is supported by the striking dataset distribution shift observed between Figures \ref{fig:protein_na_complex_len_distr} and \ref{fig:protein_na_complex_plus_monomers_len_distr}.

Table \ref{subtab:na_generation} shows that both \textsc{MMDiff} (trained on protein-nucleic acid complexes) and \textsc{MMDiff-NA} (trained solely on nucleic acid complexes) outperform the random baseline method for nucleic acid complex generation in terms of designability (i.e., $\mathrm{scRMSD}$ and $\mathrm{scTM}$). Important to note is that \textsc{MMDiff} generalizes markedly beyond the designability of the random baseline by generating a reasonable number of moderately designable, highly diverse, and novel complex structures. Figure \ref{fig:designability_scrmsd_results} illustrates the designability scores \textsc{MMDiff} can achieve in terms of $\mathrm{scRMSD}$ across a variety of complex lengths and chain compositions. Lastly, Figures \ref{fig:designed_micro_rna} and \ref{fig:designed_micro_s_dna} illustrate micro-RNA (\textsc{GCGCGCGGGG}) and single-stranded micro-DNA (\textsc{TCTTGGTTTGTCTTTTCGA}) molecules successfully designed using \textsc{MMDiff}, which achieve $\mathrm{scRMSD}$ measures of 1.26 and 4.95 and maximum trainTM measures of 0.36 and 0.31 with respect to the PDB complexes \textsc{4WTF} and \textsc{1XPX}, respectively. Moreover, RoseTTAFold2NA is confident in its predicted structure for these designed sequences, yielding per-residue average plDDT scores of \underline{95.11} and \underline{78.24}, respectively. Note, for illustrating these molecules, the corresponding \textsc{MMDiff}-generated structures are shown in green (a) or orange (b), whereas the RF2NA-predicted structures are shown in orange (a) or green (b).

Table \ref{subtab:protein_na_complex_generation} and Figure \ref{fig:designability_scrmsd_results} show that \textsc{MMDiff} outperforms the random baseline as well as \textsc{MMDiff-Monomer}, similar to the results listed in Table \ref{subtab:protein_generation} for protein generation. We argue this result is also likely due to increased modeling space complexity introduced by adding a majority of protein monomers to the training dataset of \textsc{MMDiff-Monomer}. Figure \ref{fig:designed_protein_dna_complex} illustrates a protein-DNA complex (DNA sequence: \textsc{GGGGGG}) designed by \textsc{MMDiff}. Note that, here, \textsc{MMDiff} generates a structurally-novel protein-DNA interaction in comparison to that of RF2NA while maintaining the main structure of the DNA molecule ($\mathrm{scRMSD}$ of DNA structure $< 5$ \r{A}), where RF2NA's per-residue average plDDT of 58.10 may indicate the possibility of multiple conformations for such an interaction type (as suggested by \textsc{MMDiff}). Note, for visualizing this macromolecule, the \textsc{MMDiff}-generated chains are shown in green (orange), whereas the RF2NA-predicted chains are shown in pink (purple). Also, note that all corresponding designability results in terms of $\mathrm{scTM}$ are displayed in Appendix \ref{appendix:additional_designability_results} within Figure \ref{fig:designability_sctm_results}.

\subsection{Related Work}
\label{section:related_work}
\textbf{Generative methods for molecule generation.} Several kinds of generative models have been proposed for molecular structure generation \citep{hoogeboom2022equivariant, eguchi2022ig, xu2023geometric} and sequence generation \citep{hsu2022learning}, most commonly diffusion generative models \citep{anand2022protein, wu2022protein, trippe2022diffusion, lin2023generating, fu2023latent, watson2023novo}. Regarding diffusion generative models, most closely related to this work is the $\mathrm{SE(3)}$ diffusion model of \cite{yim2023se} and the joint protein sequence-structure diffusion model of \cite{lisanza2023joint}. The former is designed solely to generate protein backbone structures, and the latter method uses a pre-trained structure prediction network to design protein sequence-structure pairs using pure Gaussian noise. In contrast to these methods, \textsc{MMDiff} designs sequence-structure pairs of not only proteins but also nucleic acid complexes using a geometrically-expressive rigid body denoising process for structure generation and a discrete diffusion process for sequence generation, thereby deriving the benefits of both data representations for macromolecular inputs while avoiding the need for pre-training.

\textbf{Nucleic acid modeling.} Few works have rigorously explored geometric modeling of nucleic acid structures to the extent that methods such as AlphaFold 2 \citep{jumper2021highly} have done for amino acid residues. Some preliminary works in this direction are those of \cite{shen2022e2efold, baek2022accurate, joshi2023multi}. Nonetheless, no prior works have presented a geometric modeling scheme for nucleic acids that is amenable to macromolecular design tasks while being complementary to amino acid modeling, which we present in this work.

\section{Discussions \& Conclusions}
\label{section:discussion}
In this work, we introduced \textsc{MMDiff}, a macromolecular sequence-structure diffusion generative model for nucleic acid and protein complexes. Through rigorous experiments within our proposed computational benchmark for macromolecular design, we have demonstrated that \textsc{MMDiff} can generate designable, diverse, and novel nucleic acid structures.

\textbf{Limitations.} Currently, \textsc{MMDiff} achieves many more success cases for nucleic acid design compared to protein(-nucleic acid) design, suggesting room for improvement with regards to joint structural modeling of both types of molecules. Our preliminary investigations into this topic suggest that acquiring new sources of diverse and high-quality macromolecular training data will be key for future developments, as the PDB is currently quite limited in the number of protein-nucleic acid quaternary structures it offers.
We argue this based on our observation that extended training of \textsc{MMDiff} on protein-nucleic acid complexes, over time, leads the model to increase its designability metrics for nucleic acids yet decrease its designability metrics for proteins and protein-nucleic acid complexes, hinting at a phenomenon akin to overfitting.

\textbf{Future work.} Given the limitations above, future work could involve (1) curating large new training datasets containing protein \textit{complexes} with or without DNA and RNA molecules for joint or separate generation tasks; (2) extending \textsc{MMDiff} to support \textit{full-atom} protein-nucleic design in an end-to-end manner; (3) validating \textsc{MMDiff}'s nucleic acid designs through wet-lab experiments; or (4) investigating efficient ways of scoring generative models producing large multi-chain molecular assemblies such as protein-RNA and protein-DNA complexes.
In this work, we have already observed the difficulties of assessing such macromolecular design methods, so we believe standardized benchmarking amongst future methods will be critical to ensure progress in the field.

% In this work, we introduced \textsc{MMDiff}, a generative model for nucleic acid and protein complex sequences and structures, which sets the stage for the development of future macromolecular generative models. Preliminary results with \textsc{MMDiff} demonstrate its ability to generate diverse and novel nucleic acid structures in computational benchmarking. However, it faces limitations in achieving success with protein complex and joint protein-nucleic acid design, suggesting a need for improved training data. As such, future work could involve (1) curating larger high-quality macromolecular datasets; (2) investigating new diffusion noise schedules; (3) validating nucleic acid designs experimentally; or (4) exploring efficient scoring methods for complex molecular assemblies.

\begin{ack}
The authors would like to thank Jason Yim for insightful discussions regarding the $\mathrm{SE(3)}$ Diffusion framework and its successful applications in protein backbone design. This work was funded by Profluent Bio and performed while Alex Morehead was a summer research intern with the company.
\end{ack}

\bibliography{neurips_2023}

\begin{thebibliography}{29}
\providecommand{\natexlab}[1]{#1}
\providecommand{\url}[1]{\texttt{#1}}
\expandafter\ifx\csname urlstyle\endcsname\relax
  \providecommand{\doi}[1]{doi: #1}\else
  \providecommand{\doi}{doi: \begingroup \urlstyle{rm}\Url}\fi

\bibitem[Anand and Achim(2022)]{anand2022protein}
Namrata Anand and Tudor Achim.
\newblock Protein structure and sequence generation with equivariant denoising diffusion probabilistic models.
\newblock \emph{arXiv preprint arXiv:2205.15019}, 2022.

\bibitem[Austin et~al.(2021)Austin, Johnson, Ho, Tarlow, and van~den Berg]{NEURIPS2021_958c5305}
Jacob Austin, Daniel~D. Johnson, Jonathan Ho, Daniel Tarlow, and Rianne van~den Berg.
\newblock Structured denoising diffusion models in discrete state-spaces.
\newblock In M.~Ranzato, A.~Beygelzimer, Y.~Dauphin, P.S. Liang, and J.~Wortman Vaughan, editors, \emph{Advances in Neural Information Processing Systems}, volume~34, pages 17981--17993. Curran Associates, Inc., 2021.
\newblock URL \url{https://proceedings.neurips.cc/paper_files/paper/2021/file/958c530554f78bcd8e97125b70e6973d-Paper.pdf}.

\bibitem[Baek et~al.(2022)Baek, McHugh, Anishchenko, Baker, and DiMaio]{baek2022accurate}
Minkyung Baek, Ryan McHugh, Ivan Anishchenko, David Baker, and Frank DiMaio.
\newblock Accurate prediction of nucleic acid and protein-nucleic acid complexes using rosettafoldna.
\newblock \emph{bioRxiv}, pages 2022--09, 2022.

\bibitem[Berman et~al.(2000)Berman, Westbrook, Feng, Gilliland, Bhat, Weissig, Shindyalov, and Bourne]{berman2000protein}
Helen~M Berman, John Westbrook, Zukang Feng, Gary Gilliland, Talapady~N Bhat, Helge Weissig, Ilya~N Shindyalov, and Philip~E Bourne.
\newblock The protein data bank.
\newblock \emph{Nucleic acids research}, 28\penalty0 (1):\penalty0 235--242, 2000.

\bibitem[Chen et~al.(2022)Chen, Zhang, and Hinton]{chen2022analog}
Ting Chen, Ruixiang Zhang, and Geoffrey Hinton.
\newblock Analog bits: Generating discrete data using diffusion models with self-conditioning.
\newblock \emph{arXiv preprint arXiv:2208.04202}, 2022.

\bibitem[Eguchi et~al.(2022)Eguchi, Choe, and Huang]{eguchi2022ig}
Raphael~R Eguchi, Christian~A Choe, and Po-Ssu Huang.
\newblock Ig-vae: Generative modeling of protein structure by direct 3d coordinate generation.
\newblock \emph{PLoS computational biology}, 18\penalty0 (6):\penalty0 e1010271, 2022.

\bibitem[Evans et~al.(2021)Evans, O’Neill, Pritzel, Antropova, Senior, Green, {\v{Z}}{\'\i}dek, Bates, Blackwell, Yim, et~al.]{evans2021protein}
Richard Evans, Michael O’Neill, Alexander Pritzel, Natasha Antropova, Andrew Senior, Tim Green, Augustin {\v{Z}}{\'\i}dek, Russ Bates, Sam Blackwell, Jason Yim, et~al.
\newblock Protein complex prediction with alphafold-multimer.
\newblock \emph{biorxiv}, pages 2021--10, 2021.

\bibitem[Fu et~al.(2023)Fu, Yan, Wang, Au, McThrow, Komikado, Maruhashi, Uchino, Qian, and Ji]{fu2023latent}
Cong Fu, Keqiang Yan, Limei Wang, Wing~Yee Au, Michael McThrow, Tao Komikado, Koji Maruhashi, Kanji Uchino, Xiaoning Qian, and Shuiwang Ji.
\newblock A latent diffusion model for protein structure generation.
\newblock \emph{arXiv preprint arXiv:2305.04120}, 2023.

\bibitem[Gelbin et~al.(1996)Gelbin, Schneider, Clowney, Hsieh, Olson, and Berman]{gelbin1996geometric}
Anke Gelbin, Bohdan Schneider, Lester Clowney, Shu-Hsin Hsieh, Wilma~K Olson, and Helen~M Berman.
\newblock Geometric parameters in nucleic acids: sugar and phosphate constituents.
\newblock \emph{Journal of the American Chemical Society}, 118\penalty0 (3):\penalty0 519--529, 1996.

\bibitem[Hie and Yang(2022)]{HIE2022145}
Brian~L. Hie and Kevin~K. Yang.
\newblock Adaptive machine learning for protein engineering.
\newblock \emph{Current Opinion in Structural Biology}, 72:\penalty0 145--152, 2022.
\newblock ISSN 0959-440X.
\newblock \doi{https://doi.org/10.1016/j.sbi.2021.11.002}.
\newblock URL \url{https://www.sciencedirect.com/science/article/pii/S0959440X21001457}.

\bibitem[Ho et~al.(2020)Ho, Jain, and Abbeel]{ho2020denoising}
Jonathan Ho, Ajay Jain, and Pieter Abbeel.
\newblock Denoising diffusion probabilistic models.
\newblock \emph{Advances in neural information processing systems}, 33:\penalty0 6840--6851, 2020.

\bibitem[Hoogeboom et~al.(2021)Hoogeboom, Nielsen, Jaini, Forr{\'e}, and Welling]{hoogeboom2021argmax}
Emiel Hoogeboom, Didrik Nielsen, Priyank Jaini, Patrick Forr{\'e}, and Max Welling.
\newblock Argmax flows and multinomial diffusion: Learning categorical distributions.
\newblock \emph{Advances in Neural Information Processing Systems}, 34:\penalty0 12454--12465, 2021.

\bibitem[Hoogeboom et~al.(2022)Hoogeboom, Satorras, Vignac, and Welling]{hoogeboom2022equivariant}
Emiel Hoogeboom, V{\i}ctor~Garcia Satorras, Cl{\'e}ment Vignac, and Max Welling.
\newblock Equivariant diffusion for molecule generation in 3d.
\newblock In \emph{International conference on machine learning}, pages 8867--8887. PMLR, 2022.

\bibitem[Hsu et~al.(2022)Hsu, Verkuil, Liu, Lin, Hie, Sercu, Lerer, and Rives]{hsu2022learning}
Chloe Hsu, Robert Verkuil, Jason Liu, Zeming Lin, Brian Hie, Tom Sercu, Adam Lerer, and Alexander Rives.
\newblock Learning inverse folding from millions of predicted structures.
\newblock In \emph{International Conference on Machine Learning}, pages 8946--8970. PMLR, 2022.

\bibitem[Jamasb et~al.(2022)Jamasb, Vi{\~n}as~Torn{\'e}, Ma, Du, Harris, Huang, Hall, Li{\'o}, and Blundell]{jamasb2022graphein}
Arian Jamasb, Ramon Vi{\~n}as~Torn{\'e}, Eric Ma, Yuanqi Du, Charles Harris, Kexin Huang, Dominic Hall, Pietro Li{\'o}, and Tom Blundell.
\newblock Graphein-a python library for geometric deep learning and network analysis on biomolecular structures and interaction networks.
\newblock \emph{Advances in Neural Information Processing Systems}, 35:\penalty0 27153--27167, 2022.

\bibitem[Joshi et~al.(2023)Joshi, Jamasb, Viñas, Harris, Mathis, and Liò]{joshi2023multi}
Chaitanya~K. Joshi, Arian~R. Jamasb, Ramon Viñas, Charles Harris, Simon Mathis, and Pietro Liò.
\newblock Multi-state rna design with geometric multi-graph neural networks.
\newblock \emph{arXiv preprint arXiv:2305.14749}, 2023.

\bibitem[Jumper et~al.(2021)Jumper, Evans, Pritzel, Green, Figurnov, Ronneberger, Tunyasuvunakool, Bates, {\v{Z}}{\'\i}dek, Potapenko, et~al.]{jumper2021highly}
John Jumper, Richard Evans, Alexander Pritzel, Tim Green, Michael Figurnov, Olaf Ronneberger, Kathryn Tunyasuvunakool, Russ Bates, Augustin {\v{Z}}{\'\i}dek, Anna Potapenko, et~al.
\newblock Highly accurate protein structure prediction with alphafold.
\newblock \emph{Nature}, 596\penalty0 (7873):\penalty0 583--589, 2021.

\bibitem[Li et~al.(2022)Li, Thickstun, Gulrajani, Liang, and Hashimoto]{li2022diffusion}
Xiang Li, John Thickstun, Ishaan Gulrajani, Percy~S Liang, and Tatsunori~B Hashimoto.
\newblock Diffusion-lm improves controllable text generation.
\newblock \emph{Advances in Neural Information Processing Systems}, 35:\penalty0 4328--4343, 2022.

\bibitem[Lin and AlQuraishi(2023)]{lin2023generating}
Yeqing Lin and Mohammed AlQuraishi.
\newblock Generating novel, designable, and diverse protein structures by equivariantly diffusing oriented residue clouds.
\newblock \emph{arXiv preprint arXiv:2301.12485}, 2023.

\bibitem[Lin et~al.(2022)Lin, Akin, Rao, Hie, Zhu, Lu, dos Santos~Costa, Fazel-Zarandi, Sercu, Candido, et~al.]{lin2022language}
Zeming Lin, Halil Akin, Roshan Rao, Brian Hie, Zhongkai Zhu, Wenting Lu, Allan dos Santos~Costa, Maryam Fazel-Zarandi, Tom Sercu, Sal Candido, et~al.
\newblock Language models of protein sequences at the scale of evolution enable accurate structure prediction.
\newblock \emph{BioRxiv}, 2022:\penalty0 500902, 2022.

\bibitem[Lisanza et~al.(2023)Lisanza, Gershon, Tipps, Arnoldt, Hendel, Sims, Li, and Baker]{lisanza2023joint}
Sidney~Lyayuga Lisanza, Jacob~Merle Gershon, Sam Wayne~Kenmore Tipps, Lucas Arnoldt, Samuel Hendel, Jeremiah~Nelson Sims, Xinting Li, and David Baker.
\newblock Joint generation of protein sequence and structure with rosettafold sequence space diffusion.
\newblock \emph{bioRxiv}, pages 2023--05, 2023.

\bibitem[Lutz(2010)]{LUTZ2010734}
Stefan Lutz.
\newblock Beyond directed evolution—semi-rational protein engineering and design.
\newblock \emph{Current Opinion in Biotechnology}, 21\penalty0 (6):\penalty0 734--743, 2010.
\newblock ISSN 0958-1669.
\newblock \doi{https://doi.org/10.1016/j.copbio.2010.08.011}.
\newblock URL \url{https://www.sciencedirect.com/science/article/pii/S0958166910001540}.
\newblock Chemical biotechnology – Pharmaceutical biotechnology.

\bibitem[Shen et~al.(2022)Shen, Hu, Peng, Chen, Xiong, Hong, Zheng, Wang, King, Wang, et~al.]{shen2022e2efold}
Tao Shen, Zhihang Hu, Zhangzhi Peng, Jiayang Chen, Peng Xiong, Liang Hong, Liangzhen Zheng, Yixuan Wang, Irwin King, Sheng Wang, et~al.
\newblock E2efold-3d: end-to-end deep learning method for accurate de novo rna 3d structure prediction.
\newblock \emph{arXiv preprint arXiv:2207.01586}, 2022.

\bibitem[Trippe et~al.(2022)Trippe, Yim, Tischer, Baker, Broderick, Barzilay, and Jaakkola]{trippe2022diffusion}
Brian~L Trippe, Jason Yim, Doug Tischer, David Baker, Tamara Broderick, Regina Barzilay, and Tommi Jaakkola.
\newblock Diffusion probabilistic modeling of protein backbones in 3d for the motif-scaffolding problem.
\newblock \emph{arXiv preprint arXiv:2206.04119}, 2022.

\bibitem[Watson et~al.(2023)Watson, Juergens, Bennett, Trippe, Yim, Eisenach, Ahern, Borst, Ragotte, Milles, et~al.]{watson2023novo}
Joseph~L Watson, David Juergens, Nathaniel~R Bennett, Brian~L Trippe, Jason Yim, Helen~E Eisenach, Woody Ahern, Andrew~J Borst, Robert~J Ragotte, Lukas~F Milles, et~al.
\newblock De novo design of protein structure and function with rfdiffusion.
\newblock \emph{Nature}, pages 1--3, 2023.

\bibitem[Wu et~al.(2022)Wu, Yang, van~den Berg, Zou, Lu, and Amini]{wu2022protein}
Kevin~Eric Wu, Kevin~K Yang, Rianne van~den Berg, James Zou, Alex~Xijie Lu, and Ava~P Amini.
\newblock Protein structure generation via folding diffusion.
\newblock \emph{arXiv preprint arXiv:2209.15611}, 2022.

\bibitem[Xu et~al.(2023)Xu, Powers, Dror, Ermon, and Leskovec]{xu2023geometric}
Minkai Xu, Alexander~S Powers, Ron~O Dror, Stefano Ermon, and Jure Leskovec.
\newblock Geometric latent diffusion models for 3d molecule generation.
\newblock In \emph{International Conference on Machine Learning}, pages 38592--38610. PMLR, 2023.

\bibitem[Yim et~al.(2023)Yim, Trippe, De~Bortoli, Mathieu, Doucet, Barzilay, and Jaakkola]{yim2023se}
Jason Yim, Brian~L Trippe, Valentin De~Bortoli, Emile Mathieu, Arnaud Doucet, Regina Barzilay, and Tommi Jaakkola.
\newblock Se (3) diffusion model with application to protein backbone generation.
\newblock \emph{arXiv preprint arXiv:2302.02277}, 2023.

\bibitem[Zhang et~al.(2022)Zhang, Shine, Pyle, and Zhang]{zhang2022us}
Chengxin Zhang, Morgan Shine, Anna~Marie Pyle, and Yang Zhang.
\newblock Us-align: universal structure alignments of proteins, nucleic acids, and macromolecular complexes.
\newblock \emph{Nature methods}, 19\penalty0 (9):\penalty0 1109--1115, 2022.

\end{thebibliography}
\bibliographystyle{plainnat}

\newpage
\section{Supplementary Material}
\subsection{Additional Designability Results}
\label{appendix:additional_designability_results}

\begin{figure}[!htb]
    \centering
    \includegraphics[width=\linewidth]{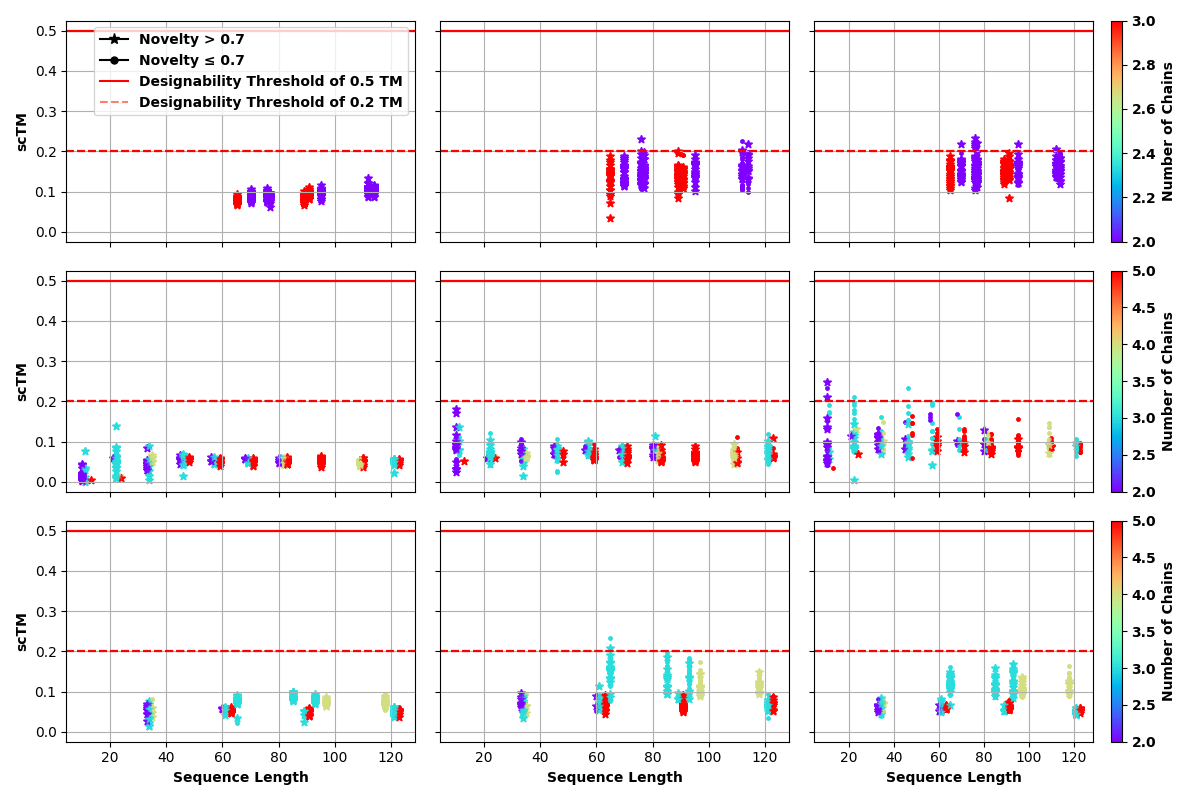}
    
    \caption{Comparison of $\mathrm{scTM}$ complex designability results using different training methods. Here, the top row corresponds to protein-only experiments, the middle row to nucleic acid-only experiments, and the bottom row to protein-nucleic acid experiments. The columns denotes samples generated using the random macromolecule generation baseline, \textsc{MMDiff}, and \textsc{MMDiff-\{Protein, NA, Monomer\}} (corresponding to the \{first, second, third\} row), respectively. Note that novel data samples are displayed with a \textsc{*} symbol. Overall, most designable complexes contain 2-3 chains, and most generated complexes contain novel chains (with a novelty $> 0.7$).}
    \label{fig:designability_sctm_results}
\end{figure}

In Figure \ref{fig:designability_sctm_results}, we report designability results in terms of $\mathrm{scTM}$ for different training methods as a complement to the results in Figure \ref{fig:designability_scrmsd_results}.

\subsection{Alternative Noise Schedules for Sequence Generation}
\label{appendix:alternative_sequence_noise}

\begin{figure}[t]
    \centering
    \begin{subfigure}[b]{0.45\linewidth}
        \includegraphics[width=\linewidth]{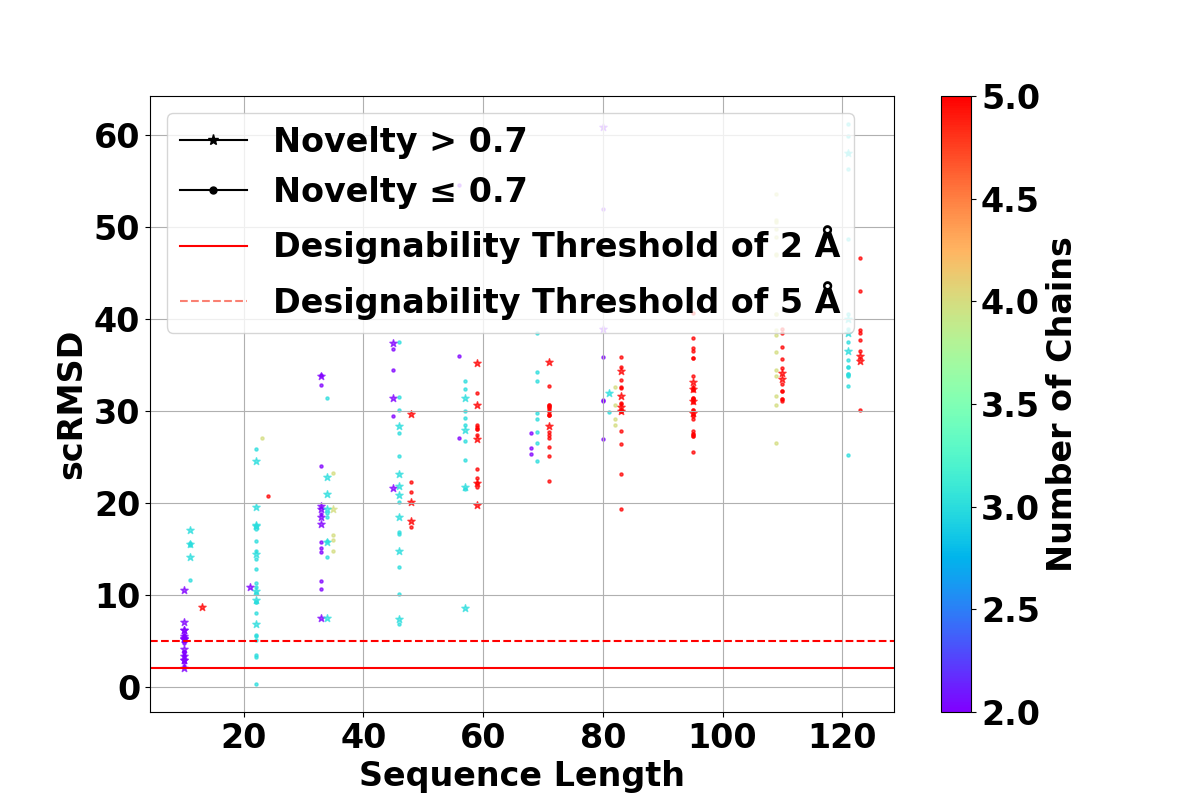}
        \caption{$\mathrm{scRMSD}$ designability results for nucleic acid structures using a version of \textsc{MMDiff} trained only on nucleic acid complexes and sampled using linear sequence noise.}
        \label{fig:appendix_na_only_specialist_linear_scrmsd_scores}
    \end{subfigure}
    \hfill
    \begin{subfigure}[b]{0.45\linewidth}
        \includegraphics[width=\linewidth]{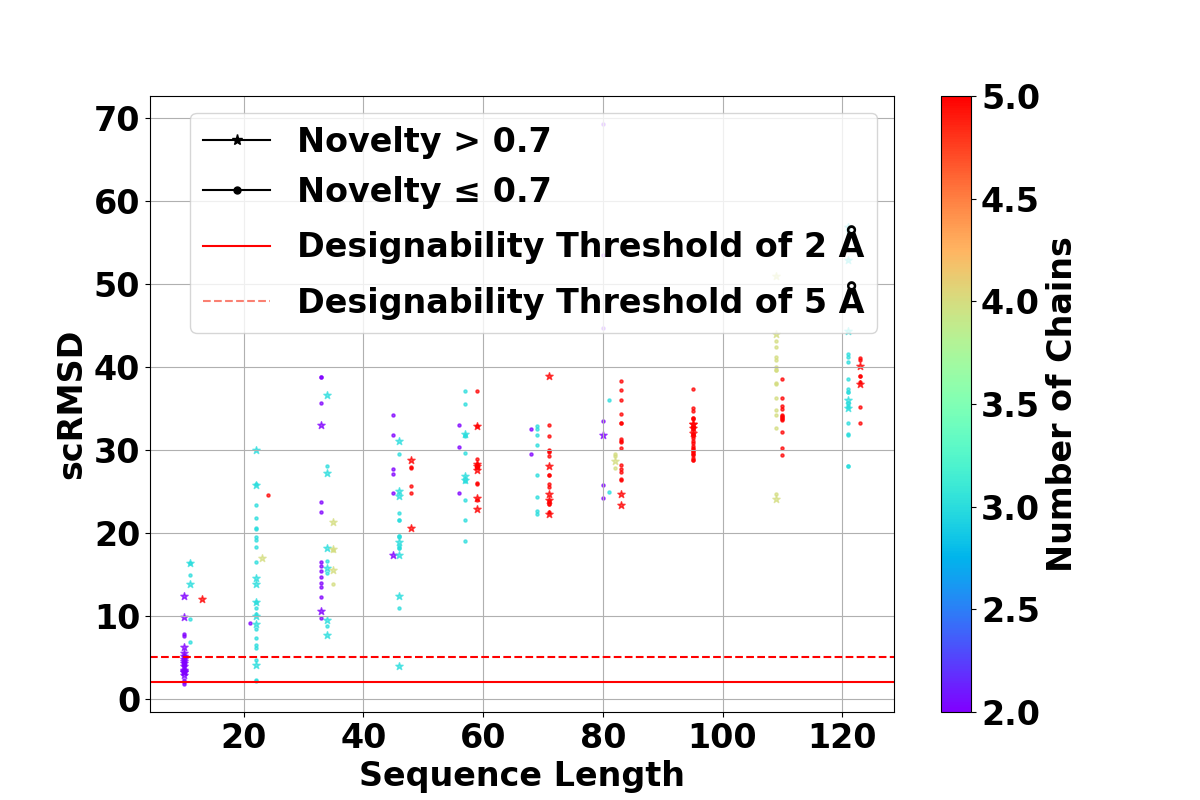}
        \caption{$\mathrm{scRMSD}$ designability results for nucleic acid structures using a version of \textsc{MMDiff} trained only on nucleic acid complexes and sampled using cosine sequence noise.}
        \label{fig:appendix_na_only_specialist_sqrt_scrmsd_scores}
    \end{subfigure}
    
    \caption{Comparison of $\mathrm{scRMSD}$ complex designability results for nucleic acid structures using different sequence noise schedules during sampling. Zoom in for the best viewing experience.}
    \label{fig:appendix_na_only_scrmsd_designability_alt_noise_schedules}
\end{figure}

\begin{figure}[t]
    \centering
    \begin{subfigure}[b]{0.45\linewidth}
        \includegraphics[width=\linewidth]{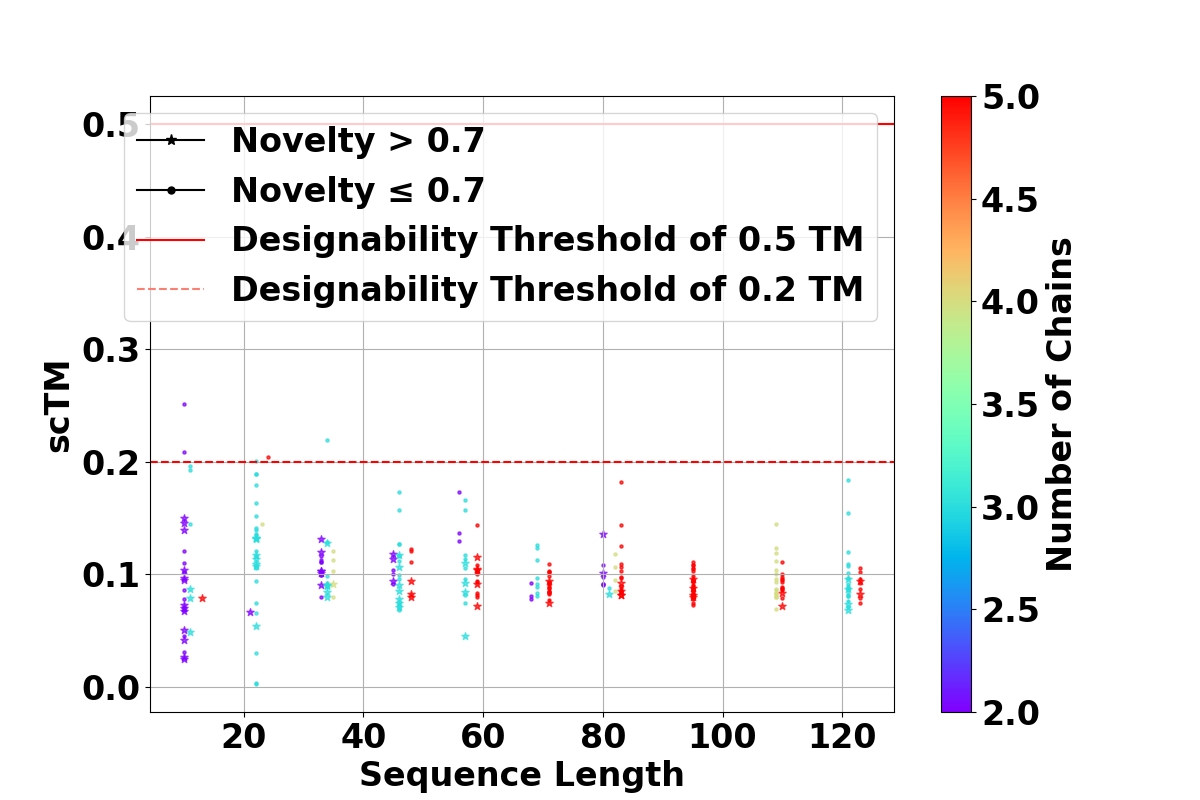}
        \caption{$\mathrm{scTM}$ designability results for nucleic acid structures using a version of \textsc{MMDiff} trained only on nucleic acid complexes and sampled using linear sequence noise.}
        \label{fig:appendix_na_only_specialist_linear_sctm_scores}
    \end{subfigure}
    \hfill
    \begin{subfigure}[b]{0.45\linewidth}
        \includegraphics[width=\linewidth]{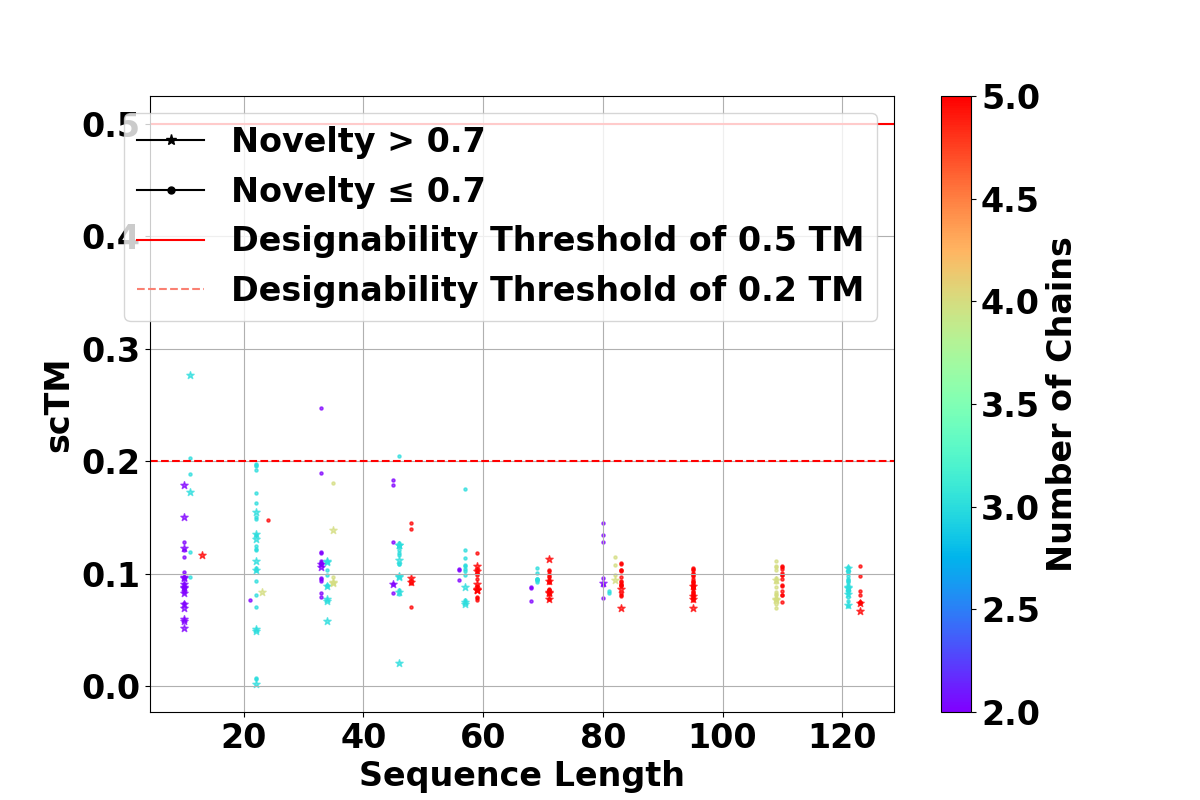}
        \caption{$\mathrm{scTM}$ designability results for nucleic acid structures using a version of \textsc{MMDiff} trained only on nucleic acid complexes and sampled using cosine sequence noise.}
        \label{fig:appendix_na_only_specialist_sqrt_sctm_scores}
    \end{subfigure}
    
    \caption{Comparison of $\mathrm{scTM}$ complex designability results for nucleic acid structures using different sequence noise schedules during sampling. Zoom in for the best viewing experience.}
    \label{fig:appendix_na_only_sctm_designability_alt_noise_schedules}
\end{figure}

As Figures \ref{fig:appendix_na_only_scrmsd_designability_alt_noise_schedules} and \ref{fig:appendix_na_only_sctm_designability_alt_noise_schedules} illustrate, in terms of designability metrics, linear and cosine sequence noise schedules marginally outperform a square root noise schedule, producing slightly more designable macromolecules overall. However, to simplify the presentation of results, in the main text we report \textsc{MMDiff} generation results using a square root noise schedule.

\newpage
\subsection{Dataset Distributions}
\label{appendix:dataset_distributions}

\begin{figure}[t]
\centering
\includegraphics[width=0.5\linewidth]{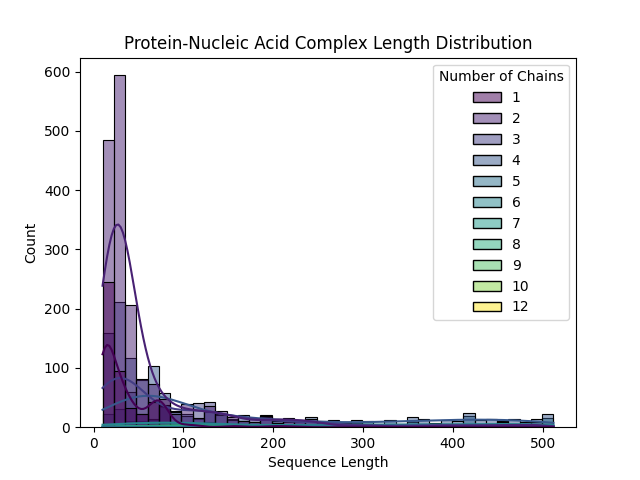}
\caption{The distribution of protein-nucleic acid complex lengths upon which \textsc{MMDiff} is trained.}
\label{fig:protein_na_complex_len_distr}
\end{figure}

\begin{figure}[t]
\centering
\includegraphics[width=0.5\linewidth]{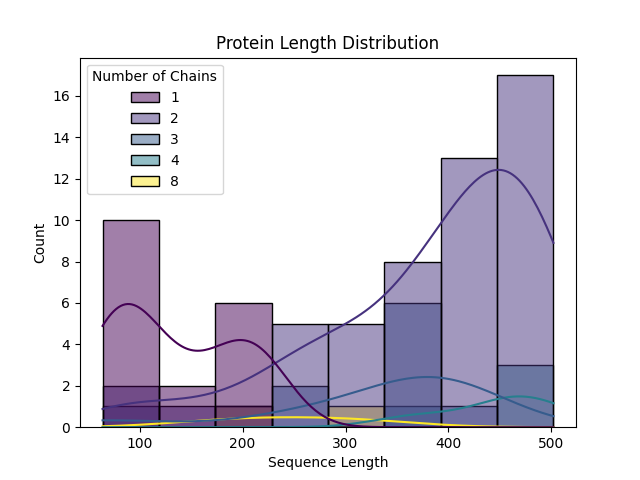}
\caption{The distribution of protein monomer and protein complex lengths upon which \textsc{MMDiff} is trained.}
\label{fig:protein_na_protein_only_len_distr}
\end{figure}

\begin{figure}[t]
\centering
\includegraphics[width=0.5\linewidth]{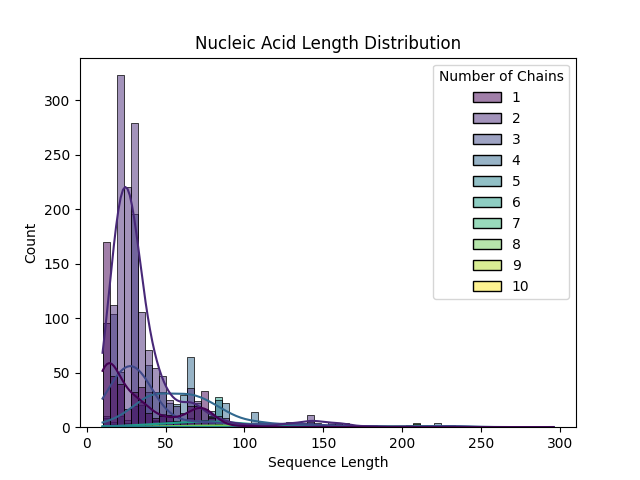}
\caption{The distribution of nucleic acid complex lengths upon which \textsc{MMDiff} is trained.}
\label{fig:protein_na_na_only_len_distr}
\end{figure}

\begin{figure}[t]
\centering
\includegraphics[width=0.5\linewidth]{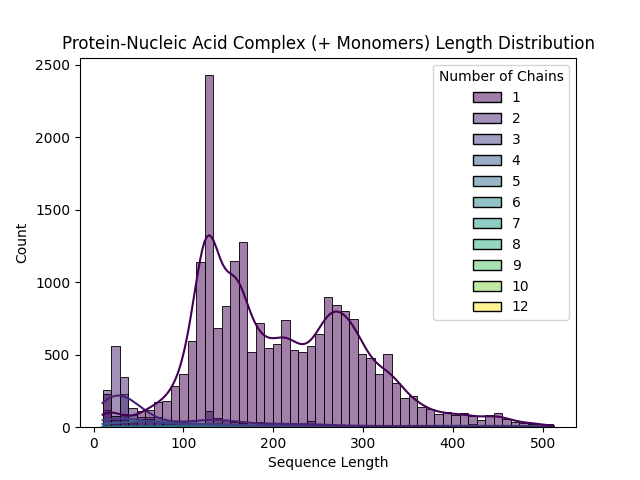}
\caption{The distribution of protein-nucleic acid complex (and monomer) lengths upon which \textsc{MMDiff-Monomer} (for protein-nucleic acid complex generation) is trained.}
\label{fig:protein_na_complex_plus_monomers_len_distr}
\end{figure}

\begin{figure}[t]
\centering
\begin{subfigure}{\linewidth}
  \centering
  \includegraphics[width=0.5\linewidth]{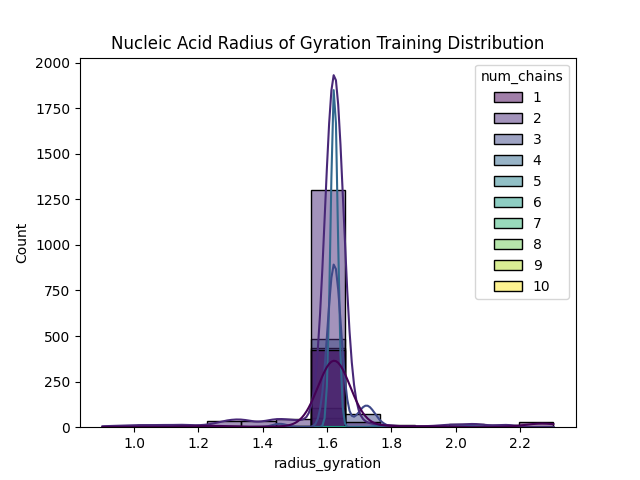}
  \caption{The radius of gyration training distribution for nucleic acid structures.}
  \label{fig:na_only_training_radius_gyration_histplot}
\end{subfigure}

\begin{subfigure}{\linewidth}
  \centering
  \includegraphics[width=0.5\linewidth]{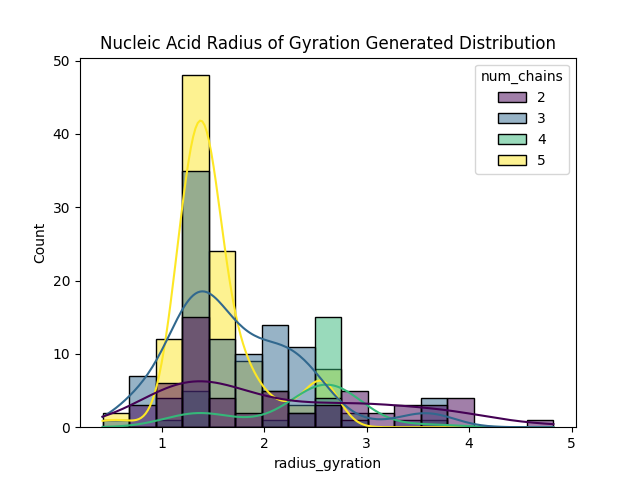}
  \caption{The radius of gyration generated distribution for nucleic acid structures.}
  \label{fig:na_only_generated_radius_gyration_histplot}
\end{subfigure}
\caption{A comparison between the radius of gyration distributions of \textsc{MMDiff}'s training dataset and \textsc{MMDiff}'s generated samples.}
\label{fig:combined_figure}
\end{figure}

Figure \ref{fig:protein_na_complex_len_distr} displays the length distribution of protein-nucleic acid complexes within \textsc{MMDiff}'s protein-nucleic acid training dataset, whereas Figures \ref{fig:protein_na_protein_only_len_distr} and \ref{fig:protein_na_na_only_len_distr} display (in isolation) the length distributions of protein complexes and nucleic acid complexes, respectively, within the protein-nucleic acid training dataset. Likewise, Figure \ref{fig:protein_na_complex_plus_monomers_len_distr} illustrates the length distribution of protein-nucleic acid complexes (and protein monomers) within \textsc{MMDiff}'s training dataset after combining it with the monomeric protein structure dataset of \cite{yim2023se}.

Lastly, to investigate how well \textsc{MMDiff} can model the true radius of gyration distribution for nucleic acid structures in its training dataset, Figures \ref{fig:na_only_training_radius_gyration_histplot} and \ref{fig:na_only_generated_radius_gyration_histplot} display the distributions of (1) \textsc{MMDiff}'s training dataset and (2) \textsc{MMDiff}'s generated samples. As shown in these figures, \textsc{MMDiff} generates its samples with an average radius of gyration of 1.8, which is reasonably close to the 1.65 average radius of gyration of its training dataset.

%%%%%%%%%%%%%%%%%%%%%%%%%%%%%%%%%%%%%%%%%%%%%%%%%%%%%%%%%%%%

\end{document}